# Sharing Rewards in Cooperative Connectivity Games

**Yoram Bachrach**                                                          YOBACH@MICROSOFT.COM
*Microsoft Research, Cambridge, UK*

**Ely Porat**                                                               PORATELY@CS.BIU.AC.IL
*Bar-Ilan University, Ramat-Gan, Israel*

**Jeffrey S. Rosenschein**                                                  JEFF@CS.HUJI.AC.IL
*The Hebrew University, Jerusalem, Israel*

## Abstract

We consider how selfish agents are likely to share revenues derived from maintaining connectivity between important network servers. We model a network where a failure of one node may disrupt communication between other nodes as a cooperative game called the vertex Connectivity Game (CG). In this game, each agent owns a vertex, and controls all the edges going to and from that vertex. A coalition of agents wins if it fully connects a certain subset of vertices in the graph, called the *primary* vertices.

Power indices measure an agent's ability to affect the outcome of the game. We show that in our domain, such indices can be used to both determine the fair share of the revenues an agent is entitled to, and identify significant possible points of failure affecting the reliability of communication in the network. We show that in general graphs, calculating the Shapley and Banzhaf power indices is #P-complete, but suggest a polynomial algorithm for calculating them in trees.

We also investigate finding stable payoff divisions of the revenues in CGs, captured by the game theoretic solution of the core, and its relaxations, the $\epsilon$-core and least core. We show a polynomial algorithm for computing the core of a CG, but show that testing whether an imputation is in the $\epsilon$-core is coNP-complete. Finally, we show that for trees, it is possible to test for $\epsilon$-core imputations in polynomial time.

## 1. Introduction

A key aspect of multi-agent systems that has been the focus of research in the field is agent collaboration. Cooperative game theory considers cooperation among self interested agents, and has been used to analyze many collaborative domains (Goldman & Zilberstein, 2004; Kraus, Shehory, & Taase, 2004; Branzei, Dimitrov, & Tijs, 2008; Dunne, van der Hoek, Kraus, & Wooldridge, 2008; Chalkiadakis, Elkind, & Wooldridge, 2012). One important application area for multi-agent systems is network analysis, examining issues ranging from communication network design (Babaoglu, Meling, & Montresor, 2002), through sensor network technologies (Lesser, Ortiz, & Tambe, 2003) to social network analysis (Sabater & Sierra, 2002). Game theory has already been used to analyze interaction between selfish agents in various network settings, such as network security (Roy, Ellis, Shiva, Dasgupta, Shandilya, & Wu, 2010; Jain, Korzhyk, Vanek, Conitzer, Pechoucek, & Tambe, 2011), resource sharing (Suris, DaSilva, Han, & MacKenzie, 2007) and agent cooperation in communication networks (Saad, Han, Debbah, Hjorungnes, & Basar, 2009; Easley & Kleinberg, 2010).

Consider a computer network in which the servers are controlled by selfish agents, and a principal interested in allowing communication between a certain set of *critical servers*. To allow this connectivity, the principal can incentivize the agents to allow communication by offering them a certain reward. These agents must then *cooperate* to allow a reliable communication between the critical servers, as every single agent only controls a small part of the entire network.

## 1.1 Key Questions and Game Theoretic Solutions

Given the above setting, several key questions arise. First, the reward is promised to the entire group of agents, who must decide how to allocate this reward amongst themselves. Even if the agents form a successful team, this team may not be *stable*, as agents who are only allocated a small part of the reward may try to form a different coalition, so as to increase their own share of the reward. Could the agents reach an agreement on sharing the rewards that would prevent such deviations? Second, what is the *fair* share that each agent should get? How can we measure the importance of each individual agent in bringing about the desired outcome of communication between all the critical servers?

Cooperative game theory provides answers to questions regarding reward sharing between selfish agents, in the form of solution concepts. Some solution concepts, such as the core and its relaxations (Gillies, 1953; Shapley & Shubik, 1966) focus on stability, whereas concepts such as the power index proposed by Banzhaf (1965) and the Shapley value (1953, 1954) focus on fairness.

Power indices originated in work on analyzing power in voting scenarios. Researchers analyzing the distribution of power in decision making bodies have tried to find a precise way of measuring the influence of a single agent in the context of a team of agents who attempt to reach a joint decision through a voting procedure. They have formalized such measures of influence as so-called *power indices*, which measure the control a voter has over decisions of a larger group (Elkind, Goldberg, Goldberg, & Wooldridge, 2007b). The two most prominent such indices are the Banzhaf power index (1965) and the Shapley-Shubik power index (1954). Each of these indices can be characterized using a set of axioms which describe desirable properties of a measure of power in voting contexts (Lehrer, 1988; Shapley, 1953; Dubey & Shapley, 1979; Straffin, 1988). The Shapley-Shubik power index is a manifestation of the Shapley value (1953) which was designed to find the *fair* share of each agent when a team of agents must cooperate to achieve a joint reward. Although these indices were mostly used for measuring power in voting systems, they can easily be adapted for other domains as well.

In this paper, we consider the use of the power indices to find fair ways for agents to share rewards in the network setting described above. Further, we show that power indices can be used to find key points of failure in a communication network. We model the above communication network setting, consisting of its servers and the network links connecting them, as a *vertex connectivity game*. The network is modeled as a graph, where the servers are the vertices, and the network links are the edges. A certain subset of the servers (vertices) are *primary*—a failure to send information between any two of them would constitute a major system failure. Another subset of the servers are always available (backbone servers).

In the vertex *Connectivity Game* (CG) that we introduce, each agent controls a different vertex in the graph. A coalition of agents can use any of the vertices controlled by the coalition members or the backbone vertices, and may send information between them. The coalition wins if it connects all the primary vertices, so that it can send information between any two of them. The power index of an agent in this game reflects its criticality in maintaining this connectivity. This index can be used to determine the fair share of the total reward that this agent should get, and could enable an administrator to identify potential critical points of failure in the network (perhaps, for example, focusing more maintenance resources on preventing their failure).

We consider the computational complexity of calculating the Banzhaf or Shapley-Shubik power indices in this domain. We show that in general graphs, computing either of these indices is a #P-complete problem. Despite this negative result, we provide a polynomial algorithm for the restricted case where the graph is a tree. Many networks, including parts of the internet's backbone, are constructed as trees when the construction of a communication line is expensive, so this algorithm can analyze important real-world domains.

We then turn to finding *stable* payoff distributions for the collaborating agents, using the game theoretic solution concept of the core (Gillies, 1953). We show that the core can be computed in polynomial time in CGs. When a coalition in the CG manages to connect all the primary vertices, it wins and gains a certain profit. This profit should then be divided among the members of the coalition. Choosing a payoff vector in the core guarantees that no subcoalition would choose to split from the main coalition, and attempt to establish its own network. Thus, the core indicates which payoff vectors are stable and allows allocating the gains of a coalition in a CG domain so as to prevent subcoalitions from defecting. We also consider the more relaxed solution concepts of the $\epsilon$-core and the least core (Shapley & Shubik, 1966), and show that testing for $\epsilon$-core imputations is coNP-complete in CGs. Under $\epsilon$-core imputations, although the coalition may not be completely stable, the incentive of any subcoalition to deviate is low. Finally, we show that in tree CGs, the core and least core coincide.

The paper proceeds as follows. In Section 2, we provide background information regarding coalitional games and power indices, and fully define a vertex connectivity game (CG). In Section 3 we examine game theoretic solutions to CGs. Section 3.1 discusses fair reward distributions in CGs using power indices, and presents a hardness result for the general case and a polynomial algorithm for the restricted case of trees. Section 3.2 examines stable reward distributions. It shows that the core of CGs can be computed in polynomial time, discusses the complexity of $\epsilon$-core and least-core-related problems, and examines core-related problems in tree CGs. Section 4 examines related work, discussing both previous work on solutions to cooperative games (Section 4.1) and related models of cooperative games over networks (Section 4.2). We conclude in Section 5.

## 2. Preliminaries

A coalitional game is composed of a set of $n$ agents, $I = (a_1, \ldots, a_n)$, and a function mapping any subset (coalition) of the agents to a real value $v : 2^I \to \mathbb{R}$. The function $v$ is called the *coalitional function* (or sometimes the *characteristic function*) of the game. In a *simple* coalitional game, $v$ only gets values of 0 or 1, so $v : 2^I \to \{0, 1\}$. A coalition $C \subseteq I$ *wins* if $v(C) = 1$, and *loses* if $v(C) = 0$. The set of all winning coalitions is denoted $W(v) = \{C \subseteq 2^I | v(C) = 1\}$. An agent $a_i$ is *critical* in a winning coalition $C$ if the agent's removal from that coalition would make the coalition lose: $v(C) = 1$ but $v(C \setminus \{i\}) = 0$. Thus, an agent can only be critical in a coalition that contains it.

We are interested in finding the fair share of the rewards an agent should get in a cooperative game, or to measuring the influence a given agent has on the result of the game. Game theoretic solutions for doing so include various values or power indices. The two most prominent such values are the Banzhaf power index (1965) and the Shapley value (1953).

Both these indices can be characterized using a slightly different sets of fairness axioms, which reflect desired properties of a power index. Both indices have the property that *dummy* agents, who never affect the value of any coalition, obtain an index of zero (the null player axiom). Similarly, both under the Shapley value and the Banzhaf index, *equivalent* agents, who increase the value of any coalition that contains neither of them by the same amount, have the same index (the symmetry axiom). However, the two indices behave differently regarding the composition of two games with the same set of agents.[1] Alternatively, these indices can be interpreted as the probability that an agent would significantly affect the outcome of the game, under slightly different models for uncertainty regarding agents' participation in the game (Straffin, 1988). Although power indices were widely used for measuring political power in weighted voting systems, their definition does not rely on the specific features of a voting domain.

The Banzhaf index depends on the number of coalitions in which an agent is critical. Agent $a_i$'s marginal contribution in a coalition $C$ where $a_i \in C$ is defined as $v(C) - v(C \setminus \{a_i\})$. Thus if $a_i$ is critical in $C$ he has a marginal contribution of 1 in it, and if $a_i$ is not critical in $C$ he has a marginal contribution of 0. The Banzhaf index of agent $a_i$ is his average marginal contribution in all coalitions that contain him , or equivalently the proportion of coalitions where $i$ is critical in out of all the coalitions that contain $i$.

**Definition 1.** *The Banzhaf index is the vector $\beta(v) = (\beta_1(v), \ldots, \beta_n(v))$ where*

$$\beta_i(v) = \frac{1}{2^{n-1}} \sum_{C \subseteq N | a_i \in C} [v(C) - v(C \setminus \{a_i\})].$$

The Shapley value (1953), which is sometimes referred to as the Shapley-Shubik power index (1954) when applied to a simple cooperative game, relies on the notion of the marginal contribution of an agent in a *permutation*. This is the amount of additional utility generated when that agent joins the coalition of her predecessors in the permutation. We denote by $\pi$ a permutation of the $n$ agents, so $\pi : \{1, \ldots, n\} \rightarrow \{1, \ldots, n\}$ and $\pi$ is onto. We denote by $S_n$ the set of all such agent permutations. Denote by $\Gamma_i^\pi$ the predecessors of $a_i$ in $\pi$, so $\Gamma_i^\pi = \{a_j | \pi(j) < \pi(i)\}$. Agent $a_i$'s marginal contribution in the permutation $\pi$ is $m_i^\pi = v(\Gamma_i^\pi \cup \{a_i\}) - v(\Gamma_i^\pi)$. Note that in a simple game an agent has a marginal contribution of 1 in the permutation $\pi$ iff it is critical for the coalition $\Gamma_i^\pi \cup \{a_i\}$. The Shapley value of an agent is her marginal contribution averaged across all possible agent permutations.

---

1. Given two games $u, v$ over the same agent set, we can define the game $u + v$ where the value of a coalition $C$ in the game $u + v$ is the sum of values in the composing games so $u + v(C) = u(C) + v(C)$. We can also define the max game $u \vee v$ where the value of a coalition in the composed game $u \vee v$ is the maximal value in the composing games, so $u \vee v(C) = max(u(C), v(C))$. Similarly we can define the min game $u \wedge v$, where $u \wedge v(C) = min(u(C), v(C))$. The Shapley value fulfills a linear decomposition axiom: the Shapley value of any agent in the sum game $u + v$ is the sum of its Shapley values in the composing games $u$ and $v$. In contrast, the Banzhaf index fulfills a property regarding max and min games, where the sum of powers in two games is the sum of the powers in their max and min games. Some related work examines the fairness axioms of the Shapley value and Banzhaf index (Dubey & Shapley, 1979; Straffin, 1988; Lehrer, 1988; Holler & Packel, 1983; Laruelle & Valenciano, 2001).

**Definition 2.** *The Shapley value is the vector $(\phi_1(v), \ldots, \phi_n(v))$ where*

$$\phi_i(v) = \frac{1}{n!} \sum_{\pi \in S_n} m_i^\pi = \frac{1}{n!} \sum_{\pi \in S_n} \left( v\left(\Gamma_i^\pi \cup \{i\}\right) - v\left(\Gamma_i^\pi\right)\right)$$

Both the Shapley value and the Banzhaf index can be thought of as the *expected* marginal contribution of an agent under certain assumptions about the coalition formation process. The Shapley value reflects the assumption that agents are randomly added to a coalition, so every *ordering* of the agents is equally probable. In contrast, the Banzhaf index reflects the assumption that all *coalitions* are equally probable. More generally, power indices can be viewed as probabilities of events in weighted voting domains (Straffin, 1988).

Our hardness result for calculating power indices in CGs considers the class #P. #P is the set of integer-valued functions that express the number of accepting computations of a nondeterministic Turing machine of polynomial time complexity. Let $\Sigma$ be the finite input and output alphabet for Turing machines.

**Definition 3.** *#P is the class consisting of the functions $f : \Sigma^* \rightarrow \mathbb{N}$ such that there exists a non-deterministic polynomial time Turing machine $M$ that for all inputs $x \in \Sigma^*$, $f(x)$ is the number of accepting paths of $M$.*

The complexity classes #P and #P-complete were introduced by Valiant (1979a). These classes express the hardness of problems that "count the number of solutions".[2]

The coalitional function $v$ describes the total utility a coalition can achieve, but does not define how the agents distribute this utility among themselves. An *imputation* $(p_1, \ldots, p_n)$, sometimes also called a *payoff vector*, is a division of the gains of the grand coalition $I$ among the agents, where $p_i \in \mathbb{R}$, and $\sum_{i=1}^n p_i = v(I)$. We call $p_i$ the payoff of agent $a_i$, and denote the payoff of a coalition $C$ as $p(C) = \sum_{i \in \{j | a_j \in C\}} p_i$. By assumption, agents are rational and attempt to maximize their own share of the utility. Game theory offers several solution concepts, determining which imputations are likely to occur when agents act rationally.

The Shapley value has been shown to be an imputation, as the values of the individual agents were shown to sum up to the value of the grand coalition of all the agents: $\sum_{i=1}^n \phi_i(v) = v(I)$.[3] Given the fairness axioms that the Shapley value fulfills, it can thus be viewed as a fair imputation. However, in many domains the agents are selfish and care little for fairness. Rather, selfish agents are likely to be more interested in their ability to improve their own utility by forming alternative coalitions. Stability based solution concepts such as the core focus on such deviations (Gillies, 1953).

---

2. Informally, NP and NP-hardness deal with checking if at least one solution to a combinatorial problem exists, while #P and #P-hardness deal with calculating the *number* of solutions to a combinatorial problem. Counting the number of solutions to a problem is at least as hard as determining if there is at least one solution, so #P-complete problems are at least as hard (but possibly harder) than NP-complete problems; These complexity classes have been thoroughly investigated by computation complexity researchers (Papadimitriou, 2003; Valiant, 1979b; Papadimitriou & Zachos, 1982).

3. Shapley provided a proof of this fact in his seminal paper on the Shapley value (1953).

A basic stability requirement for an imputation is *individual rationality*: for all agents $a_i \in C$, we have $p_i \geq v(\{a_i\})$, otherwise that agent is better off on its own. Similarly, we say that a coalition $B$ *blocks* the imputation $(p_1, \ldots, p_n)$ if $p(B) < v(B)$, since the members of $B$ would split off from the coalition and gain more utility without the rest of the agents. If a blocked imputation is chosen, the coalition is unstable. It is possible to define the *degree* by which a subcoalition is incentivized to deviate from the grand coalition.

**Definition 4.** *Given an imputation $p = (p_1, \ldots, p_n)$, the excess of a coalition is $e(C) = v(C) - p(C)$, which quantifies the amount the subcoalition $C$ can gain by deviating and working on its own.*

Given an imputation, a coalition $C$ is blocking iff its excess is strictly positive $e(C) > 0$. If a blocked payoff vector is chosen, the coalition is unstable, and the higher the excess is, the more incentivized the agents are to split apart from the current coalition and form their own coalition. A known solution concept that emphasizes stability is the core (Gillies, 1953).

**Definition 5.** *The Core of a coalitional game is the set of all payment vectors $(p_1, \ldots, p_n)$ that are not blocked by any coalition, so for any coalition $C$ we have $p(C) \geq v(C)$.*

A value distribution in the core makes sure that no subset of the agents would split off, so the coalition is stable. In general the core can be empty, so every possible value division is blocked by some coalition. In this paper, we give results regarding computing the core in vertex connectivity games. When the core can be empty (i.e., every possible value division is blocked by some coalition), it sometimes make sense to relax the requirements of the solution concept. In some domains, splitting apart from the current coalition structure to form an alternative coalition might be a costly process. In such cases, coalitions that only have a small incentive to split apart from the grand coalition would not do so. A relaxed solution concept embodying this intuition is the $\epsilon$-core (Shapley & Shubik, 1966). The $\epsilon$-core slightly relaxes the inequalities of Definition 5.

**Definition 6.** *The $\epsilon$-core is the set of all imputations $(p_1, \ldots, p_n)$ such that the following holds: for any coalition $C \subseteq I$, $p(C) \geq v(C) - \epsilon$.*

Under an imputation in the $\epsilon$-core, the excess $e(C) = v(C) - p(C)$ of any coalition $C$ is at most $\epsilon$. For large enough values of $\epsilon$, the $\epsilon$-core is guaranteed to be non-empty. An obvious problem is to find the smallest value of $\epsilon$ that makes the $\epsilon$-core non-empty. This solution concept is known as the *least core*. Formally, consider the game $G$ and the set $\{\epsilon | \text{the } \epsilon\text{-core of G is not empty}\}$. This set is compact,[4] so it has a minimal element $\epsilon_{min}$.

**Definition 7.** *The least core of the game $G$ is the $\epsilon_{min}$-core of $G$.*

---

4. A formal a formal definition of compactness and its implications can be found in many introductory books on topology (Royden & Fitzpatrick, 1988).

Imputations in the least core distribute payoffs while minimizing the worst deficit. In other words, the least core minimizes the maximal incentive of a coalition to split apart from the grand coalition. Under a least core imputation, no coalition can gain more than $\epsilon_{min}$ by deviating, and for any $\epsilon' < \epsilon_{min}$ it is impossible to distribute the payoffs in a way that causes the deficit of any coalition to be at most $\epsilon'$. Another solution concept, called the nucleolus (Schmeidler, 1969) refines the least core, minimizing the *number* of coalitions that have the maximal excess by examining the sorted vector of excesses and defining a lexicographical order over them.

## 2.1 Connectivity Games

Consider a network connecting various servers, where a certain subset of the servers are designated "primary" servers. Our goal is to make sure that we can send information between any two primary servers. A server in the network may malfunction, and if it does, we cannot send information through it. If all the paths between two primary servers go through a failed server, we cannot send information between these two primary servers. In our model, we also assume that there can be a certain subset of servers that are guaranteed never to fail (guaranteed, say, by heavy maintenance and fail-safe backup); we will call these "backbone" servers.

In Section 1.1 we discussed several questions regarding agreements selfish agents are likely to reach when each of them controls a server in such a network. We model the network domain as a cooperative game and use game theoretic solutions to answer these questions. The coalitional game at the heart of our model is called the vertex *Connectivity Game*.

**Definition 8.** *A vertex Connectivity Game Domain (CGD) consists of a graph $G = \langle V, E \rangle$ where the vertices are partitioned into primary vertices $V_p \subseteq V$, backbone vertices $V_b \subseteq V$, and standard vertices $V_s \subseteq V$. We require that $V_p \cap V_b = \emptyset$, $V_b \cap V_s = \emptyset$, $V_p \cap V_s = \emptyset$, and that $V = V_p \cup V_b \cup V_s$, so this is indeed a partition.*

Given a CGD, we can define the vertex Connectivity Game. In this game, each agent controls one of the standard servers. A coalition wins if it connects all pairs of primary vertices (so it can send information between any two such primary servers). Let $|V_s| = n$, and consider a set of $n$ agents $I = (a_1, \ldots, a_n)$, so that agent $a_i$ controls vertex $v_i \in V_s$. Given a coalition $C \subseteq I$ we denote the set of vertices that $C$ controls as $V(C) = \{v_i \in V_s | a_i \in C\}$. Coalition $C$ can use either the vertices in $V(C)$ or the always-available backbone vertices $V_b$. In our model, we assume that the coalition can also use any of the primary vertices $V_p$ as well.[5]

We say a set of vertices $V' \subseteq V$ fully connects $V_p$ if for any two vertices $u, v \in V_p$ there is a path $(u, p_1, p_2, \ldots, p_k, v)$ from $u$ to $v$ going only through vertices in $V'$, so for all $i$ we have $p_i \in V'$.

---

5. Another possibility would be to allow some of the primary vertices we want to connect to fail (this may occur, for example, in network security domains, where external attacks may target key servers in the network). In this case a coalition would win if it manages to connect all the non-failed primary vertices. We could also disallow sending information through the primary vertices (so they can only be the final destination). Most of the results in this paper hold for these different settings as well.

**Definition 9.** *A vertex Connectivity Game (CG) is a simple coalitional game, where the value of a coalition $C \subseteq I$ is defined as follows:*

$$v(C) = \begin{cases} 1 & \text{if } V(C) \cup V_b \cup V_p \text{ fully connects } V_p \\ 0 & \text{otherwise} \end{cases}$$

## 3. Solutions To The Connectivity Game

In Section 1.1 we raised several questions regarding agreements selfish agents are likely to reach when each of them controls a server in such a network. We now answer these questions by applying game theoretic solutions on our Connectivity Game.

We begin by characterizing fair payoff distribution and measuring an agent's importance in allowing reliable network communication. Specifically, given our desire to ensure communication paths between "primary" vertices, which servers on the network are most critical? Which agents deserve a higher share of the reward? Given limited resources to make sure that some servers do not fail (i.e., making them backbone servers), on which vertices should we concentrate to ensure communication between primary servers? Section 3.1 answers these questions using power indices.

We continue by examining stable allocations of the rewards. How can we determine whether an agreement on sharing the rewards would incentivize a sub-coalition of agents to defect? If no agreement is fully-resistant to such deviations, how can we find the most stable agreement? Section 3.2.1 answers these questions using core and core-related solutions.

### 3.1 Network Reliability And Fair Reward Distribution

The CG of Definition 9 has a characteristic function that maps every coalition that fully connects the primary vertices to a single unit of reward (and indicates that all other coalitions have a reward of zero). In Section 2 we have discussed the fairness axioms that the Shapley value and the Banzhaf index fulfill. Due to these properties, we can apply these concepts to CGs, and obtain a fair distribution of the reward in these games. For example, dummy servers, which do not affect the ability of any coalition to allow full connectivity, would have a power index of zero, and obtain no reward. Similarly, equivalent agents, that have the same impact on achieving connectivity when added to any coalition, would obtain the same reward. If the agents forming the coalition wish to share the rewards in a "fair" manner, power indices thus make an excellent basis for forming the agreement,[6] highlighting the need to examine the computational aspects of calculating such indices, as we do in this section.

We emphasize that beyond fulfilling these fairness properties, power indices can also be viewed as network reliability constructs. Our model is based on a simple network goal of allowing communication between any two primary servers. Under a network reliability view, we want to identify the servers which, when failing, will cause us to lose connectivity between primary servers. Suppose all the servers have an equal probability of working or failing the next day (i.e. each has a probability of 50% to fail, and a probability of 50% to work). When these failures are independent, any subset of the servers has an equal chance of surviving (regardless of its size). Thus, we have a certain probability of having the surviving set of servers fully connect the primary servers. Suppose we can make sure that exactly one server, owned by agent $a_i$, always survives. The Banzhaf power index measures the increase in the probability of having the surviving subset of vertices fully connect the primary servers by guaranteeing that $v_i$ survives.

---

6. The specific index to be used depends on the agents' notion of fairness. Different sets of axioms result in different indices (Dubey & Shapley, 1979; Straffin, 1988; Holler & Packel, 1983; Laruelle & Valenciano, 2001).

When attempting to maximize the probability of achieving our goal, the higher the Banzhaf index of a server is, the more we should try to make sure that server does not fail. Thus, in order to find significant points of failure, we can calculate the Banzhaf power index, and focus on the servers with the highest indices.[7]

We now consider the computational complexity of calculating power indices in general vertex connectivity games. We first formally define the problems.

**Definition 10.** *CG-BANZHAF / CG-SHAPLEY: We are given a CG over the graph $G = \langle V, E \rangle$, with primary vertices $V_p \subseteq V$, backbone vertices $V_b \subseteq V$, and standard vertices $V_s \subseteq V$. There are $n = |V_s|$ agents, $I = (a_1, \ldots, a_n)$, so agent $a_i$ controls vertex $v_i \in V_s$. The game's coalitional function $v : 2^I \to \{0,1\}$ is defined as in Definition 9. We are also given a specific target agent $a_i$. In the CG-BANZHAF we are asked to calculate its Banzhaf power index in this game, $\beta_i(v)$. Similarly, in the CG-SHAPLEY problem we are asked to calculate its Shapley value, $\phi_i(v)$.*

We now show that in general CGs, both CG-BANZHAF and CG-SHAPLEY are #P-complete. We first prove the problems are in #P. We then reduce a #SET-COVER problem to CG-BANZHAF. Then we obtain the #P-hardness of CG-SHAPLEY as a corollary. We begin with a few definitions.

**Definition 11.** *#SET-COVER (#SC): We are given a collection $C = \{S_1, \ldots, S_n\}$ of subsets. We denote $\cup_{S_i \in C} S_i = S$. A set cover is a subset $C' \subseteq C$ such that $\cup_{S_i \in C'} = S$. We are asked to compute the number of covers of $S$.*

A slightly different version requires finding the number of set covers *of size at most $k$*:

**Definition 12.** *#SET-COVER-K (#SC-K): A set-cover with size $k$ is a set cover $C'$ such $|C'| = k$. As in Definition 11, we are given $S$ and $C$ and a target size $k$, and are asked to compute the number of covers of $S$ of size at most $k$.*

Both #SC and #SC-K are #P-hard. Garey and Johnson (1979) showed that #SC-K is #P-hard: they considered several basic NP-complete problems, and showed that their counting versions are #P-complete. The counting version of SET-COVER discussed there is #SC-K. #VERTEX-COVER is a restricted form of #SC. Vadhan (2002) showed that #VERTEX-COVER is #P-hard,[8] so #SC is also #P-hard. We use #SC to prove that CG-BANZHAF is #P-hard. It is easy to show that #SC-K is #P-complete, but the fact that #SC is #P-complete is more difficult to prove (and is thus not very well known). We give the definitions of both #SC and #SC-K to avoid confusion between them, and use Vadhan's result (2002) which indicates that #SC is #P-complete. Using this, by reducing #SC to CG-BANZHAF we show that CG-BANZHAF is #P-complete.

---

7. Similarly, when there is uncertainty about the *order* of agent failures, the Shapley value reflects the importance of vertices with regard to network reliability.

8. He also showed that the problem remains #P-hard even in very restricted classes of graphs.

In order to show that CG-BANZHAF is #P-complete we need to show two things: first, that CG-BANZHAF is *in* #P, and second, a reduction of a #P-hard problem to CG-BANZHAF.

**Lemma 1.** *CG-BANZHAF and CG-SHAPLEY are in #P.*

*Proof.* The Banzhaf index of $a_i$ in a CG $v$ is $\beta_i(v)$, the proportion of coalitions where $a_i$ is critical, out of all the coalitions that contain $a_i$. Given a certain coalition $C \subseteq I$, it is polynomial to check whether it wins—we only need to check whether $V(C) \cup V_b$ fully connects $V_p$. We can do this by creating a new graph $G'$, dropping all edges that miss $V(C) \cup V_b$ from $G$ (i.e., we drop any edge $(x, y) \in E$ such that either $x \notin V(C) \cup V_b$ or $y \notin V(C) \cup V_b$). We then check if any two primary vertices in $G'$ are connected (there are several polynomial algorithms to do this; a simple one is to run a depth-first search (DFS) between all pairs of primary vertices). We can thus easily test if a certain agent $a_i$ is critical for a coalition: we perform the above test when he is in the coalition, remove him, and repeat the test. If the first test succeeds and the second fails, that agent is critical for that coalition. Similarly, we can test whether an agent $a_i$ is critical in an agent permutation $\pi$ in polynomial time: we simply test whether $a_i$'s predecessors $\Gamma_i^\pi$ form a losing coalition and whether $\Gamma_i^\pi\{a_i\}$ form a winning coalition (we do both using the above test).

Since we can construct a deterministic polynomial Turing machine $M$ that tests if an agent is critical in a coalition, we can construct a non-deterministic Turing machine $M'$, that first non-deterministically chooses a coalition that $a_i$ is a member of, and then tests if $a_i$ is critical in that coalition. The number of accepting paths of $M'$ is the number of coalitions that contain $a_i$ where $a_i$ is critical. Denote by $k$ the number of such accepting paths of $M'$, and denote $|I| = |V_s| = n$. Then the Banzhaf power index of agent $a_i$ is $\beta_i(v) = \frac{k}{2^{n-1}}$.

Calculating the numerator of $\beta_i(v)$ is thus, according to Definition 3, a problem in #P. Since the denominator is constant (given a domain with $n$ agents), CG-BANZHAF is in #P. A similar argument using a non-deterministic generation of permutations instead of coalitions holds for the Shapley value, so CG-SHAPLEY is in #P. $\qquad \square$

We now show that CG-BANZHAF is #P-hard. We do this by a reduction from #SC. Figure 1 shows an example of such a reduction for a specific #SC instance.

**Theorem 1.** *CG-BANZHAF is #P-hard, even if there are no backbone vertices, i.e., $V_b = \emptyset$.*

*Proof.* We reduce a #SC instance to a CG-BANZHAF instance. Consider the #SC instance with the collection $C = \{S_1, \ldots, S_n\}$, so that $\cup_{S_i \in C} S_i = S$. Denote the items in $S$ as $S = \{t_1, t_2, \ldots, t_k\}$. Denote the items in $S_i$ as $S_i = \{t_{(S_i,1)}, t_{(S_i,2)}, \ldots, t_{(S_i,k_i)}\}$. The reduction-generated CGD is constructed with a graph $G = \langle V, E \rangle$ as follows. For each subset $S_i \in C$, the generated CG instance has a vertex $v_{S_i}$. We denote the set of all such vertices as $V_{sets} = \cup_{\{i|S_i \in C\}} v_{S_i}$. For each item $t_i \in S$ the generated CG instance also has a vertex $v_{t_i}$.

We denote the set of $v_{t_i}$ vertices $V_{items} = \cup_{i|t_i \in S} v_{t_i}$. The generated CG instance also has two special vertices $v_a$ and $v_b$. These are all the vertices of the generated instance.

The vertices in the generated CG are connected in the following way. The vertices $V_{sets}$ are a clique: for every $v_i, v_j \in V_{sets}$, $(v_i, v_j) \in E$. The vertex $v_a$ is also a part of that clique, so for all $v_i \in V_{sets}$ we have $(v_i, v_a) \in E$. The vertex $v_a$ is connected to $v_b$, and is the only vertex connected to $v_b$, so $(v_a, v_b) \in E$. Each set vertex $v_{S_i}$ is connected to all the vertices of the items in that set, $v_{t_{(S_i,1)}}, v_{t_{(S_i,2)}}, \ldots, v_{t_{(S_i,k_i)}}$, so for any $v_{S_i} \in V_{sets}$ and any $v_{t_{(S_i,j)}}$ (so that $t_{(S_i,j)} \in S_i$) we have $(v_{S_i}, v_{t_{(S_i,j)}}) \in E$.

We define the CG so that $V_p = V_{items} \cup \{v_b\}$, $V_b = \emptyset$, $V_s = V_{sets} \cup \{v_a\}$, and the CG game is defined as in Definition 9. The game has $m = |V_s| = |V_{sets}| + 1 = |C| + 1 = n + 1$ agents (where $n$ is the number of subsets in $C$, the input to the #SC problem). The CG-BANZHAF query is regarding $v_a$. Let $\beta_i(v)$ be the answer to the CG-BANZHAF query, and $k$ be the number of set covers in the #SC instance. We show that $k = \beta_{v_a}(v) \cdot 2^{m-1}$, by providing a one-to-one mapping between a set-cover of the original problem and a winning coalition where $v_a$ is critical in the generated CG.

Consider a set-cover $C' \subseteq C$ for $S$. $C'$ must cover all the items $t_i$ in $S$. We denote the set of vertices corresponding to the sets in this vertex cover $V_{C'} = \{v_{S_i} \in V_{sets} | S_i \in C'\}$. Since $C'$ is a set cover for the original problem, each vertex $v_{t_j} \in V_{items}$ in the generated graph must be connected to at least one vertex $v_i \in V_{sets}$. Since the vertices $V_{sets}$ are a clique, in the generated CG all the $v_{t_i}$'s and $v_{S_j}$'s are in the same connected component. However, without $v_a$ we cannot reach $v_b$ from any vertex. Thus, $V_{C'} \cup \{v_a\} \subseteq V_S$ is a winning coalition in the generated CG, but $V_{C'}$ is not, so $v_a$ is critical for that coalition. We now show the mapping in the reverse direction. Consider a coalition $V' \subseteq V_s$ where $v_a$ is critical, and denote $C' = \{S_i \in C | v_{S_i} \in V'\}$. By definition, $V'$ must be winning and contain $v_a$. Consider any vertex $v_{t_i} \in V_{items}$. Since $V'$ wins, it must allow any vertex in $V_{sets}$ to reach $v_{t_i}$, which can only happen if $V'$ contains some $v_{S_j}$ so that $t_i \in S_j$. Thus, $C'$ is a set cover for the original problem.

Let $x$ be the number of set covers in the #SC instance, and $c_a$ be the number of winning coalitions where $v_a$ is critical in the generated CG. Due to the one-to-one mapping we have shown, $x = c_a$. But by the definition of the Banzhaf index, in the generated CG we have $\beta_a(v) = \frac{c_a}{2^{m-1}}$, so $c_a = \beta_a(v) \cdot 2^{m-1}$, and then $x = \beta_a(v) \cdot 2^{m-1}$.

We have shown that given a polynomial algorithm for CG-BANZHAF, we can solve #SC in polynomial time, so CG-BANZHAF is #P-hard. □

A recent result shows that for any *reasonable* representation language of a cooperative game, if computing the Banzhaf index is #P-hard, then computing the Shapley value is also #P-hard (Aziz, Lachish, Paterson, & Savani, 2009). A representation language is said to be reasonable if it is possible to represent the game with an additional dummy agent using the same representation language.[9] Our CG representation is reasonable—to add an additional dummy agent we simply add a dummy vertex $x$ which is not connected to any other vertex. Adding this isolated vertex to any coalition does not change its value. Thus we obtain the following corollary.

---

9. More formally, a representation language is reasonable if for any game $v$ that it can represent, it can also represent the game $v'$ defined as follows. The game $v'$ has an additional agent $x$ that is not present in the original game $v$. For any agent set $C$ such that $x \notin C$ we have $v'(C) = v(C)$. For any agent set $C$ such that $x \in C$ we have $v'(C) = v(C \setminus \{x\})$. Note that $v$ and $v'$ are games with a slightly different agent sets: $v$ is defined over an agent set $I$ whereas $v'$ is defined over the agent set $I \cup \{x\}$.

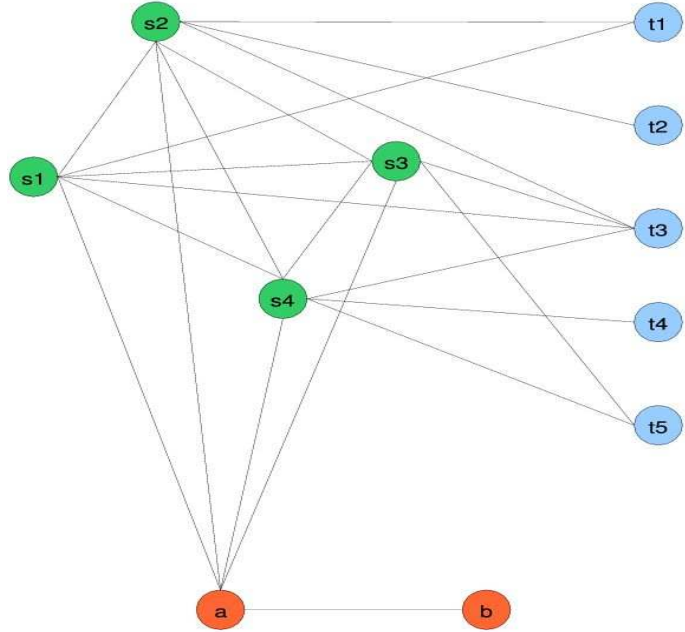

Figure 1: Example of reducing #SC to CG-BANZHAF. The items are $\{t_1, t_2, t_3, t_4, t_5\}$ and the sets are $S_1 = \{t_1, t_3\}$, $S_2 = \{t_1, t_2, t_3\}$, $S_3 = \{t_3, t_5\}$, $S_4 = \{t_3, t_4, t_5\}$.

**Corollary 1.** *CG-SHAPLEY is #P-hard, even if there are no backbone vertices, i.e., $V_b = \emptyset$.*

Since CG-BANZHAF and CG-SHAPLEY are both in #P and #P-hard these are #P-complete problems, so it is unlikely that a polynomial algorithm for calculating these indices in CGs would be found. We can circumvent this computational problem in several ways. One is to try to find an approximation algorithm, and the other is to solve the problem for restricted instances. In the next section, we adopt the second approach.

### 3.1.1 Computing Power Indices In Tree CGs

Although computing the Shapley and Banzhaf power indices in general CGs is #P-complete, restricting the graph's structure may allow us to polynomially compute such indices. We examine the restricted case where the graph is a tree. Consider a CG with graph $G = \langle V, E \rangle$ *that is a tree*, with primary vertices $V_p \subseteq V$, backbone vertices $V_b \subseteq V$, and standard vertices $V_s \subseteq V$. We call the problem of calculating the Shapley and Banzhaf power index of an agent in this domain *TREE-CG-SHAPLEY / TREE-CG-BANZHAF*. We assume that there are at least two primary vertices $v_a, v_b \in V_p$ (otherwise, any subset of the vertices trivially fully connects the primary vertices). We first note that since the graph is a tree, some of the vertices are "veto agents" present in any winning coalition.

**Lemma 2.** *Consider a CG where the graph $G$ is a tree. Let $v_a, v_b \in V_p$ be two primary vertices, and a standard vertex $v_r \in V_s$ is on a simple path from $v_a$ to $v_b$. Then $v_r$ is present in* all *winning coalitions in the CG game.*

*Proof.* Since $G$ is a tree, there is only one simple path between $v_a$ and $v_b$. The removal of any vertex along that simple path makes $v_b$ unreachable from $v_a$. Since $v_r$ is such a vertex, any coalition $C \subset V_s$ such that $v_r \notin C$ loses, and any winning coalition must contain $v_r$. □

Due to Lemma 2 we call a standard vertex in a tree-CG *essential* if it lies on the simple path between some two primary vertices.

**Lemma 3.** *Consider a CG where the graph $G$ is a tree. Let $v_a, v_b \in V_p$ be two primary vertices. Consider a vertex coalition $C \subset V_s$ that contains all standard vertices $v_r \in V_s$ on the single simple path from $v_a$ to $v_b$. Then $C$ allows the connecting of $v_a$ and $v_b$ in the CG game. If the coalition $C \subset V_s$ contains* all *essential vertices (i.e., all standard vertices $v_r \in V_s$ on any simple path between any two primary vertices $v_a, v_b \in V_p$), then $C$ is a winning coalition so $v(C) = 1$.*

*Proof.* In the CG game we can use any primary vertex $v_p \in V_p$, and any backbone vertex $v_b \in V_b$. Consider a coalition $C$ that contains all vertices $v_r \in V$ on the single simple path from $v_a$ to $v_b$. Any vertex $v_x$ on the single simple path between $v_a$ and $v_b$ is either a backbone vertex (so $v_x \in V_b$) or a primary vertex (so $v_x \in V_p$) or a standard vertex (so $v_x \in V_s$). If it is a standard vertex, it is in the coalition, so $v_x \in C$. In any of these cases we can use the vertex, so $v_a$ and $v_b$ are in the same connected component for the coalition $C$. If $C$ contains *all* essential vertices, then any two primary vertices $v_a$ and $v_b$ are connected through a path composed of either backbone, primary, or coalition vertices, so $C$ fully connects all the primary vertices $V_p$, and is a winning coalition. □

Lemma 2 states that having all essential vertices is a necessary condition for a coalition to win, and Lemma 3 states that this is also a sufficient condition. We summarize this in the following corollary.

**Corollary 2.** *In Tree CGs, the winning coalitions are exactly those coalitions that contain all essential vertices.*

We denote the set of all essential vertices in a tree CG as $V_{es}$. We show that in a tree CG the Shapley value distributes all the reward to the essential vertices equally, and provide a similar result for the Banzhaf index.

**Theorem 2.** *Consider a tree CG $v$ with a set $V_{es}$ of essential vertices. If $v_i \in V_{es}$ then $\phi_i(v) = \frac{1}{|V_{es}|}$ and otherwise $\phi_i(v) = 0$.*

*Proof.* Three of the fairness axioms the Shapley value fulfills are the null player axiom, the symmetry axiom and the efficiency axiom (Shapley, 1953; Dubey & Shapley, 1979; Straffin, 1988). The null player axiom states that if $i$ is a null player, i.e., for any coalition $C$ we have $v(C) = v(C \cup \{i\})$ then $\phi_i(v) = 0$. Due to corollary 2 if a vertex $i$ is not essential then it is a null player, so $\phi_i(v) = 0$. The symmetry axiom deals with equivalent agents. Two agents are $i, j$ are called equivalent if for any coalition $C$ that contains neither of them (so $i \notin C$ and $j \notin C$) adding either agent results in the same change in value, so $v(C \cup \{i\}) = v(C \cup \{j\})$. We note that all essential vertices are equivalent—if there exist two essential vertices, $i, j \in V_{es}$, then any coalition $C$ that does not contain either of them is missing two essential vertices to be winning, so $v(C \cup \{i\}) = v(C \cup \{j\}) = 0$. Due to the symmetry axiom, all essential vertices have the same Shapley value, which we denote $\phi_{es}$. The efficiency axiom states that the Shapley values of all the agents sum up to the value of the grand coalition $I$, so $\sum_{i \in I} \phi_i(v) = v(I)$. Due to the efficiency and null player axioms and since the Shapley value of any vertex $i \in V_{es}$ is $\phi_{es}$ we have $1 = v(I) = \sum_{i \in I} \phi_i(v) = \sum_{i \in V_{es}} \phi_i + \sum_{i \notin V_{es}} \phi_i = |V_{es}| \cdot \phi_{es} + \sum_{i \notin V_{es}} 0$, so we have: $\phi_{es} = \frac{1}{|V_{es}|}$. $\qquad \blacksquare$

**Theorem 3.** *Consider a tree CG $v$ with a set $V_{es}$ of essential vertices. If $i \in V_{es}$ then $\beta_i(v) = 2^{1-|V_{es}|}$ and otherwise $\phi_i(v) = 0$.*

*Proof.* Due to corollary 2 if a vertex $i$ is not essential then it is a null player, so for any coalition $C$ we have $v(C) - v(C \setminus \{i\}) = 0$. In this case, directly from Definition 1 we have $\beta_i(v) = 0$. Now suppose $i$ is an essential vertex. The entire agent set has $n$ vertices, where $|V_{es}|$ are essential and $n - |V_{es}|$ are non-essential. Due to Corollary 2 the winning coalitions are exactly those containing all the essential vertices, including $i$. Thus $i$ is critical in all the winning coalitions. There are therefore $2^{n-|V_{es}|}$ different winning coalitions that $i$ is a critical in (any of the $n - |V_{es}|$ non-essential vertices can can either be present or not), so we have $\beta_i(v) = \frac{2^{n-|V_{es}|}}{2^{n-1}} = 2^{n-|V_{es}|-n+1} = 2^{1-|V_{es}|}$. $\qquad \blacksquare$

**Corollary 3.** *TREE-CG-SHAPLEY and TREE-CG-BANZHAF are in P.*

*Proof.* Due to Theorems 2 and 3 to compute the power index of a vertex we need only know whether it is essential and the total number $|V_{es}|$ of essential vertices. It is easy to test if a vertex is essential in polynomial time, by checking whether it lies between two primary vertices (for example using a DFS). We can apply this test to each of the vertices and determine whether $i$ is essential and obtain $|V_{es}|$, then apply the formulas of Theorem 2 or Theorem 3 (depending on the power index we are interested in). $\qquad \blacksquare$

Thus, despite the high complexity result for the general case given in Section 3.1, in tree CGs we can polynomially calculate power indices. This result is important for analyzing reliability in real-world networks. As an example, consider the situation where Internet connectivity is established between companies, where one company is the supplier and another company is the client. An example of a cycle in this relationship would be if company A buys an Internet connection from company B, which in turn buys an Internet connection from company C, which eventually buys an Internet connection from company A. This would mean that, in a sense, company A would have become a client of itself, and would be paying money for its own connection. This scenario which has a cycle is very unlikely, so such a domain is likely to be a tree domain.

Yet another example is agent based smart-grid technology, where agents can negotiate power supply (Massoud Amin & Wollenberg, 2005; Vytelingum, Voice, Ramchurn, Rogers, & Jennings, 2010; Pipattanasomporn, Feroze, & Rahman, 2009). In this scenario various agents are both suppliers and consumers of electricity. For example, each of several firms can have its own solar panels for producing electricity, but could buy additional electricity when its demand is higher than its own production capability. In such a case, firms benefit from being able to send and receive power from all other firms (perhaps through an intermediary). Having multiple paths to transmit power between two firms does not offer any advantage, so cycles are not likely to exist, making the network a tree.

## 3.2 Stable Reward Distributions

The key question regarding CGs raised in Section 1.1 was how agents are likely to share the reward in a CG. Section 3.1 focused on *fair* allocations, according to an agent's impact on the entire coalition achieving its goal. In this section we focus on *stable* reward allocations. Once a winning coalition is formed, a reward distribution may collapse if a subset of agents who are only allocated a small share of the reward defect and form an alternative coalition. This subset of agents would only defect if by doing so it can secure its agents a larger share of the rewards. This reasoning is captured by the *core* (see Definition 5 in Section 2). Thus, computing the core allows us to find or test for stable agreements—when the core is non-empty, it contains imputations that are stable; when it is empty, the coalition would be unstable no matter how we divide the utility among the agents. However, how can we compute the core of CG?

We first note that it is not always possible to concisely represent the core, since it may contain an infinite number of imputations. However, in the case of CGs, there does exist a concise representation for the core.

Definition 9 of CGs clearly indicate that CGs are simple cooperative games, as the value of a coalition is either 1 or 0. The core is a very demanding concept in simple games. An agent $a_i$ is a *veto player* if it is present in all winning coalitions, so if $a_i \notin C$ we have $v(C) = 0$. In Section 3.1.1 we noted that essential vertices, which lie on the *only* simple path between two primary vertices, are veto players. It is a well-known fact that in simple coalitional games, the core is non-empty iff there is at least one veto player in the game (Chalkiadakis et al., 2012). Consider a simple coalitional game that has no veto players, so for every agent $a_i$ we have a winning coalition $C$ that does not contain $a_i$. Take a payoff vector $p = (p_1, \ldots, p_n)$ where $p_i > 0$. Since $\sum_{i=0}^{n} p_i = 1$ and since $p_i > 0$ we know that $p(C) \leq \sum_{p_j \in I_{-a_i}} p_j < 1$, so $p(C) < v(C) = 1$, which makes $C$ a blocking coalition. On the other hand, we can see that any payoff vector $p$ where non-veto players get nothing is in the core: any coalition $C$ that can potentially block $p$ must have $v(C) = 1$ (if $v(C) = 0$ then it cannot block), and must contain all the veto players, so $\sum_{p_j \in C} p_j = 1$, and thus cannot block $p$.

Due to the above characterization of the core in simple cooperative games, in such games the core can be represented as a set $I_{veto}$, consisting of all the veto players in that game. This set represents all core imputations: an imputation $p = (p_1, \ldots, p_n)$ is in the core if $\sum_{i \in I_{veto}} p_i = 1$ (note that it must be the case that $\sum_{i \in I} p_i = 1$ for $p$ to be an imputation).

We now consider computing the core in CGs, in the above representation as the set of veto agents. We note that CGs are monotone games. Let $W \subseteq I$ be a winning coalition in a CG (so $v(W) = 1$), and let $C \subseteq I$ be any coalition in that game. Then $W \cup C$ is also a winning coalition, so $v(W \cup C) = 1$ (this can be restated as: for all coalitions $A, B \subseteq I$ in a CG we have $v(A \cup B) \geq v(A)$). The reason for this is that if $C$ fully connects $V_p$ then $W \cup C$ also fully connects $V_p$, as more vertices are available for us to use.

We now denote the set of all the agents except $a_i$ as $I_{-i} = I \setminus \{a_i\}$. Let $G$ be the CG graph. We denote by $G_{-i}$ the same graph when we drop the vertex $v_i$ owned by $a_i$, so $G_{-i} = \langle V_{-i}, E_{-i} \rangle$ where $V_{-i} = V \setminus \{v_i\}$ and $E_{-i} = \{(u, v) \in E | u \neq v_i \wedge v \neq v_i\}$. We now show a polynomial algorithm for testing if a player is a veto agent in CGs.

**Lemma 4.** *Testing if agent $a_i$ is a veto agent in a CG is in P.*

*Proof.* We first show that $I_{-i}$ is a losing coalition iff $a_i$ is a veto agent. If $I_{-i}$ is a losing coalition then due to the monotonicity of CGs any sub-coalition of it, $C \subseteq I_{-i}$, is also losing. Thus, any coalition without $a_i$ is losing, so $a_i$ is a veto player. On the other hand, if $I_{-i}$ is a winning coalition, it is a winning coalition where $a_i$ is not present, so by definition $a_i$ is not a veto player. Thus, to test if $a_i$ is a veto agent we only need to test if $I_{-i}$ is losing or winning. According to Definition 9 of the CG, to check if $I_{-i}$ wins we need to check if $I_{-i}$ fully connects the primary vertices. This test can be performed in polynomial time by trying all pairs $v_a, v_b \in V_p$, and performing a DFS between $v_a$ and $v_b$ in the graph $G_{-i}$. $\square$

Since computing the core in simple coalitional games just requires returning a list of all the veto agents, we get the following corollary.

**Corollary 4.** *It is possible to compute and return a concise representation of the core of a CG in polynomial time. Under this representation, it is possible to test whether the core is empty or test if an imputation is in the core in polynomial time.*[10]

*Proof.* Computing the core of a CG requires returning $I_{veto}$, the set consisting of veto players in the game. Using Lemma 4, we can check all the agents to determine which of them are veto players. If there are no veto players, the core is empty. Otherwise, any payoff vector that distributes 1 (the total utility $v(I) = 1$) among the veto players and gives none to the non-veto players is in the core. $\blacksquare$

### 3.2.1 THE $\epsilon$-CORE AND LEAST CORE

The core of CGs may be non-empty, in which case we can easily compute core imputations. However, many real world networks have redundancy in terms of connectivity, and it might be possible to connect the primary vertices even after eliminating an arbitrary single vertex. In those networks, the CG has no veto agent and the core is empty. In such cases, any imputation would be unstable, as some vertex subset would be incentivized to deviate and form its own coalition. Thus, we may simply wish to *minimize* the incentive of any agent subset to deviate, and examine problems related to the *least core*. Although core-related problems in CGs can be solved in polynomial time (as we have shown above), we now show that problems related to the $\epsilon$-core may be hard.

Given a certain proposed imputation $p = (p_1, \ldots, p_n)$, we may wish to test whether this imputation is *in* the $\epsilon$-core, for a given $\epsilon$.

---

10. In fact, the same can be done for any simple monotone coalitional game where the value of a coalition can be computed in polynomial time: due to the same proof of Lemma 4, in such games we can test whether an agent is a veto player, and in simple games computing the core simply requires finding out who the veto players are.

**Definition 13.** $\epsilon$-*CORE-MEMBERSHIP(ECM): Given an imputation $p = (p_1, \ldots, p_n)$, decide whether it is in the $\epsilon$-core of the game, or in other words, test whether any coalition $C$ has an excess $e(C)$ of at most $\epsilon$.*

ECM tests whether an imputation (payoff division) is "sufficiently stable", or $\epsilon$-stable. By definition of the $\epsilon$-core, such imputations have the property that any coalition $C$ has an excess $e(C) < \epsilon$. ECM is a more basic question than *computing* an imputation in the $\epsilon$-core or finding the the least core value—the minimal $\epsilon$ that admits a non-empty $\epsilon$-core. We show that ECM is coNP-complete, but that in can be solved in polynomial time in tree CGs. For tree CGs, we show that the core is non-empty (so the least core coincides with the core) and that it is possible to find $\epsilon$-core imputations in polynomial time.

We show that ECM is coNP-complete in CGs using a reduction from VERTEX-COVER, known to be NP-complete (Garey & Johnson, 1979).

**Theorem 4.** *ECM is coNP-complete in general CGs, even if the imputation to be tested is the equal imputation $p = (\frac{1}{n}, \ldots, \frac{1}{n})$ and if there is a single backbone vertex.*

*Proof.* ECM requires testing whether for a given imputation $p = (p_1, \ldots, p_n)$ there does not exist a coalition $C$ with excess $e(C)$ of at least $d$ (for a given $d$). Given a vertex subset $C \subseteq V$, it is easy to test if $C$ connects all the primary vertices in polynomial time, and thus test whether $C$ is winning or not. We can also easily compute the payoff $p(C)$ of the coalition under the imputation, and thus can also compute its excess $e(C) = v(C) - p(C)$. Thus, ECM is *in* coNP.

We now show that computing the maximal excess $e_{max} = \max\{e(C) | C \subset I\}$ under the imputation $p$ is coNP-hard. Note that an imputation $p$ is in the $\epsilon$-core iff the maximal deficit under this imputation is at most $\epsilon$. We show that testing whether the maximal excess is at most $d$ (for a given $d$) is NP-hard by reducing a VERTEX-COVER instance to this problem.

Let the graph $G = \langle V, E \rangle$ and threshold $t$ be the input to the VERTEX-COVER instance (i.e., we are asked wether $G$'s edges could be covered by at most $t$ vertices). We will assume the VERTEX-COVER instance has at least two edges (otherwise, the problem is easy to solve). Denote $|V| = n$. We construct a graph $G' = \langle V', E' \rangle$ as follows. For each vertex $v \in V$, we create a *standard* vertex $v \in V'$ (i.e., $V \subseteq V'$). For each edge $e \in E$, we create a *primary* vertex $v_e \in V'$. These primary vertices are called the edge vertices. We also create a single *backbone* vertex $v_b \in V'$. Thus we have $V_s = V$, $V_p = \{v_e | e \in E\}$ and $V_b = \{v_b\}$. The agents are the standard vertices $V_s = V$, so there are $n$ agents.

For each edge $e = (u, v) \in V$ we create two edges in $G'$: $e_1^{(u,v)} = (u, v_e)$ and $e_2^{(u,v)} = (v, v_e)$. In other words, we "break" each edge of the original graph into two parts, putting a vertex in between. The original edges of the graph $G$ are eliminated from $G'$. Finally, we connect any standard vertex $v \in V_s = V$ to $v_b$. Figure 2 shows an example of the reduction construction used in the proof of Theorem 4.

The imputation to be tested is the equal imputation $p = (\frac{1}{n}, \ldots, \frac{1}{n})$ (i.e., the payoff of all the agents is the same, $\frac{1}{n}$). The threshold value $\epsilon$ for the generated ECM instance is $\epsilon = 1 - \frac{t}{n}$ where $t$ is the threshold in the VERTEX-COVER instance.

We first note that any coalition $C$ that loses in the generated CG (i.e., fails to connect the primary vertices $V_p$) has a negative excess, as $v(C) = 0$ and $p(C) = \frac{|C|}{n}$, so $e(C) = v(C) -$ $p(C) = -\frac{|C|}{n}$. Thus, the maximal excess coalition is the minimally paid winning coalition, i.e., a coalition $C$ that minimizes $p(C)$ of all winning coalitions $C$: $\arg\min_{C \in \{C|v(C)=1\}} p(C)$. Since $p(C) = \frac{|C|}{n}$, a maximal excess coalition is a winning coalition of a minimal size: $\arg\min_{C \in \{C|v(C)=1\}} |C|$. In other words, a maximal excess coalition is a minimally sized coalition that connects all the primary vertices.[11] A winning coalition $C$ of size $|C| = s$ thus has an excess of $v(C) - p(C) = 1 - |C| \cdot \frac{1}{n} = 1 - \frac{s}{n}$.

Since the agents in the game are the standard vertices, and since $V_s = V$, we identify the agents with the vertices of the original graph. We show that a coalition $C \subset V_s$ is a winning coalition iff it is a vertex cover in $G$.

If $C \subset V_s$ wins, it must connect any two primary vertices $v_x, v_y \in V_p$. We note that due to our construction no two primary vertices are connected directly, as each primary vertex $v_e$ was created for an edge $e \in E$ of the original graph. We have assumed the VERTEX-COVER instance has at least two edges, so there are at least two primary vertices in the generated graph. Let $v_e$ be some primary vertex. Due to our construction $v_e$ is connected to exactly two standard vertices $u, w$ (the vertices that were connected through the edge $e$ in the original graph). If neither $u$ nor $w$ are part of $C$ (i.e., both $u \notin C$ and $w \notin C$), there is no path from $v_e$ to any other vertex in the graph induced by $C$, so $v_e$ is not connected to any other primary vertex and $C$ loses so $v(C) = 0$. Thus, if $C$ is a winning coalition then for every primary vertex $v_e$ where $e = (u, w) \in E$ for vertices $u, w \in V$ in the original graph $G$, the coalition $C$ must contain either $u$ or $w$. However, if $u \in C$ then $C$ covers the edge $e$, as $u$ is a vertex on one side of the edge, and if $w \in C$ then $C$ also covers edge $e$, as $w$ is the vertex on the other side of the edge. Thus, $C$ covers any edge $e \in E$, so it is a vertex cover.

On the other hand, suppose the coalition $C \subseteq V_s = V$ is a vertex cover of $G$. As a vertex cover, $C$ must cover every edge $e \in E$, so given an edge $e = (u, w)$, $C$ must contain either $u$ or $w$ or both. If $u \in C$ we have a path from $v_e$ to $v_b$: $(v_e, v_u, v_b)$ ($v_e$ is the source primary vertex, $v_u$ is in the coalition, and $v_b$ is a backbone vertex). Similarly, if $w \in C$ we have a path from $v_e$ to $v_b$: $(v_e, v_w, v_b)$ ($v_e$ is the source primary vertex, $v_w$ is in the coalition, and $v_b$ is a backbone vertex). Therefore, there is a path from any primary vertex $v_p$ to $v_b$. Thus, all of the primary vertices are connected: given $v_x, v_y \in V_p$ we have a path from $v_x$ to $v_b$ and from $v_b$ to $v_y$, so if $C$ is a vertex cover, it is a winning coalition.

We have shown that $C$ is a winning coalition and has an excess of $1 - \frac{|C|}{n}$ in the generated instance iff it is a vertex cover of size $|C|$. The maximal excess problem in the generated instance requires finding a minimal size winning coalition, or in other words finding a vertex cover of minimal size. We can restate this by saying that the $ECM$ instance (with $\epsilon = 1 - \frac{t}{n}$) is a "yes" instance iff the original graph has a vertex cover of size at most $t$. $\square$

### 3.2.2 The Core, $\epsilon$-Core And Least Core In Tree CGs

We consider core related problems in tree CGs. A CG domain with less than two primary vertices is a degenerate domain (where all coalitions win), and a CG where even the grand coalition of all the standard vertices fails to connect all the primary vertices is also a degen-erate domain (where all coalitions lose). We assume that the CG domains in this section are not degenerate. We also assume that no two primary vertices are *directly* connected to each other or through a path containing only backbone vertices. If this is the case, we can "merge" these primary vertices to a single vertex, as any coalition that is connected to one of them is also connected to the other one without using any standard vertex.

---

11. Finding the minimally paid winning coalition under a general imputation is very similar to the famous Steiner tree problem, which is known to be NP-hard. However, in our domain the weights are on the vertices rather than the edges.

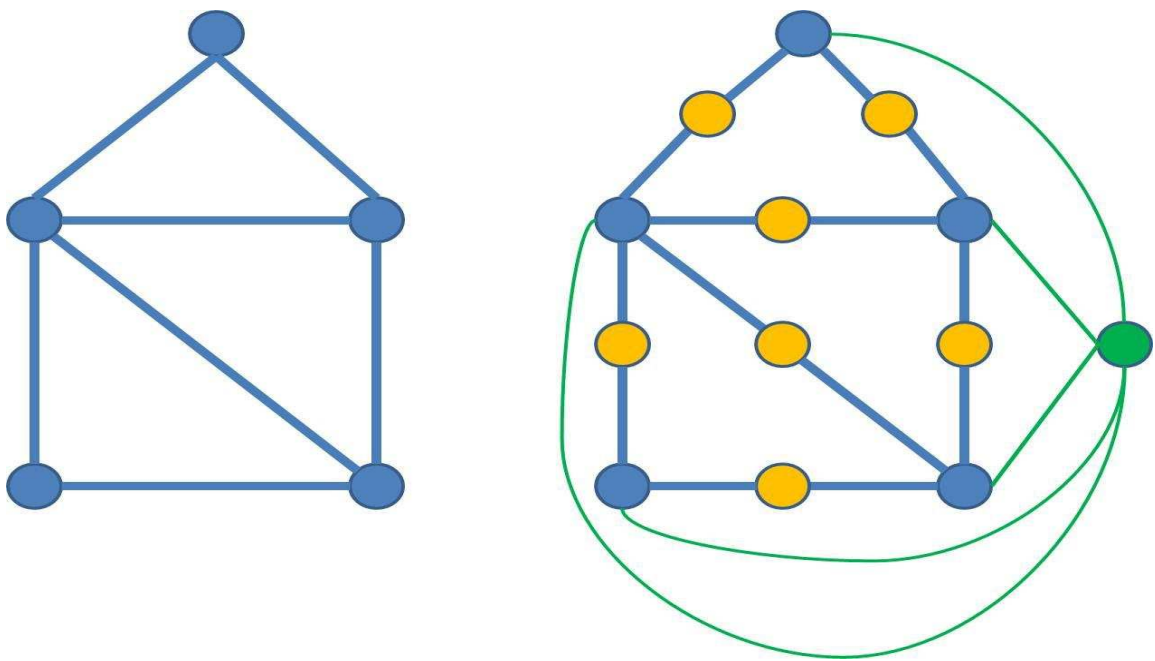

Figure 2: Example of reducing VERTEX-COVER to ECM in a CG domain. The left side is the original VERTEX-COVER instance, and the right side is the generated ECM domain, with the primary vertices in orange (denoting the edges in the original VERTEX-COVER instance), standard vertices in blue (same as the original vertices in the VERTEX-COVER instance) and the backbone vertex in green.

We first prove that the core in (non-degenerate) tree CGs is non-empty.

**Theorem 5.** *Tree CGs have non-empty cores (assuming the domain of the game is a non-degenerate CG domain).*

*Proof.* Lemma 2 Corollary 2 shows that in tree CGs, the veto vertices are exactly those essential vertices, that lie on a simple path between two primary vertices. We note that if there is no standard vertex that is on some path between two primary vertices, the domain is degenerate (either all primary vertices are connected even for the empty coalition of standard vertices, or they are still disconnected even for the grand coalition of all standard vertices). Thus, there must exist at least one veto agent. Since in simple games the core is non-empty iff there are veto agents, the core of tree CGs is non-empty. □

Due to Theorem 5, in tree CGs the least core value, the minimal $\epsilon$ such that the $\epsilon$-core is non-empty, is 0 (the core is the 0-core, and it is always non-empty).

In simple games, when the core is non-empty, any core imputation distributes the reward solely to the veto agents. Since in tree CGs the veto agents are exactly the essential vertices (Corollary 2) we obtain the following:

**Corollary 5.** *In a tree CG, let $V_{es} \subset V_s$ be the set of essential vertices, and denote $|V_{es}| = m$. Any imputation where $\sum_{i \in V_{es}} p_i = 1$ is a core imputation, and these are the only core imputations. Specifically, the imputation where $p_i = \frac{1}{m}$ if $v_i \in V_{es}$ and $p_i = 0$ otherwise is a core imputation.*

Although the core is non-empty in tree CGs, given a potential agreement in the form of a *specific* imputation, we may still wish to find the maximal excess under that imputation, or in other words test whether the imputation is in the $\epsilon$-core (for a given $\epsilon$). We now examine the complexity of the ECM problem in tree CGs. Theorem 4 has shown that this problem is coNP-complete for general graphs, but we show the problem can be solved in polynomial time for tree CGs.

**Theorem 6.** *In tree CGs, the ECM problem can be solved in polynomial time. An imputation is in the $\epsilon$-core iff $p(V_{es}) > 1 - \epsilon$.*

*Proof.* Consider a tree CG and an imputation $p = (p_1, \ldots, p_n)$. Let the veto vertices be $V_{es} \subset V$, and denote $m = |V_{es}|$. Denote the indices of the veto vertices as $M$, so $\{v_i | i \in M\} = V_{es}$. As seen in Lemma 2, in tree CGs, if a vertex $w$ is not a veto vertex (essential), it is a *dummy* vertex, so for any coalition $C$ we have $v(C \cup \{w\}) = v(C)$. We denote the non-veto vertices as $V_d = V \setminus V_{es}$, and these are all dummy vertices. We denote the indices of the dummy vertices as $D$ so $\{v_i | i \in D\} = V_d$.

We denote the total payoff of the veto vertices as $p(V_{es}) = \sum_{i \in M} p_i$, and the total payoff of the dummy vertices as $p(V_d) = \sum_{i \in D} p_i$. We assumed that the domain is not degenerate, so the grand coalition wins, and we have $v(I) = 1$. Since $p$ is an imputation, $p(I) = 1$. Since in a tree CG all vertices are either veto vertices or dummy vertices, we have $1 = p(I) = p(V_{es}) + p(V_d)$. The $\epsilon$-core constraints require that $p(C) > v(C) - \epsilon$. So since all $p_i$'s are positive, the $\epsilon$-core constraint holds for all losing coalitions. Any winning coalition $C$ must contain $V_{es}$ so $V_{es} \subseteq C$, so if $p(V_{es}) > v(V_{es}) - \epsilon = 1 - \epsilon$, *all* the $\epsilon$-core constrains hold: when $p(V_{es}) > v(V_{es}) - \epsilon = 1 - \epsilon$, for any winning coalition $C$ we have $p(C) \geq p(V_{es})$ so $p(C) \geq p(V_{es}) > v(V_{es}) - \epsilon = 1 - \epsilon = v(C) - \epsilon$. On the other hand, if we have $p(V_{es}) < 1 - \epsilon$, then the $\epsilon$-core constraint does not hold for the coalition $V_{es}$ so the imputation is not in the $\epsilon$-core.

Thus, to test if an imputation is in the $\epsilon$-core we only need to test if $p(V_{es}) > 1 - \epsilon$. Since we can compute $V_{es}$ in polynomial time, this test can also be done in polynomial time, so ECM is in P for tree CGs. $\square$

## 4. Related Work

In this paper we introduced a cooperative game called the Connectivity Game, and examined computational aspects of calculating power indices or finding core solutions in this game. We discuss related work regarding the solution concepts in Section 4.1, and examine similar models of cooperative games over networks in Section 4.2.

## 4.1 Solutions To Cooperative Games

The stability based solution concept of the core originated in the paper (Gillies, 1953). The least core was introduced as a solution concept for games with empty cores (Shapley & Shubik, 1966). The further refinement of the nucleolus was by Schmeidler (1969). The $\epsilon$-core and nucleolus were studied in minimum cost spanning tree games (Granot & Huberman, 1984), which are somewhat reminiscent of our model (see below), in assignment games (Solymosi & Raghavan, 1994), and in weighted voting games (Elkind, Goldberg, Goldberg, & Wooldridge, 2007a). Another related problem is the Cost of Stability (Bachrach, Elkind, Meir, Pasechnik, Zuckerman, Rothe, & Rosenschein, 2009), measuring the required external subsidy to stabilize a game, studied in weighted voting games (Bachrach, Meir, Zuckerman, Rothe, & Rosenschein, 2009), network flow games (Resnick, Bachrach, Meir, & Rosenschein, 2009) and other game forms (Meir, Zick, & Rosenschein, 2012; Meir, Bachrach, & Rosenschein, 2010).

Power indices originated in work on game theory and political science, attempting to measure the power that players have in weighted voting games. In these games, each player has a certain weight, and a coalition's weight is the sum of the weights of its participants; a coalition wins if its weight passes a certain threshold. This is a common situation in legislative bodies. Power indices have been suggested as a way of measuring the influence that players in such games have on choosing outcomes. The most popular indices suggested for such measurement are the Banzhaf index (1965) and the Shapley-Shubik index (1954).

The Shapley-Shubik index (1954) is a direct application of the Shapley value (1953) to simple coalitional games. The Banzhaf index emerged from the study of voting in decision-making bodies, where a certain *normalized* form of the index was introduced (Banzhaf, 1965). The Banzhaf index was later mathematically analyzed (Dubey & Shapley, 1979), and this normalization was shown to have certain undesirable features, focusing attention on the non-normalized version of the Banzhaf index. These indices were applied to an analysis of the voting structures of the IMF and the European Union Council of Ministers, as well as many other bodies (Leech, 2002; Machover & Felsenthal, 2001).

The Shapley value is the only payoff division rule that exhibits natural fairness axioms (Shapley, 1953; Dubey & Shapley, 1979), so it has been used not only to measure power but also to fairly allocate costs, revenues or credit in various domains (Shubik, 1962; Dubey, 1982; Young, 1985; Bachrach, 2010; Bachrach, Graepel, Kasneci, Kosinski, & Van Gael, 2012a; Staum, 2012). Specific such examples include dividing the costs of multicast transmissions (Feigenbaum, Papadimitriou, & Shenker, 2001), dividing airport landing fees (Littlechild & Owen, 1973), pollution reduction costs (Petrosjan & Zaccour, 2003), sharing supply chain profits (Shi-hua & Peng, 2006; Bachrach, Zuckerman, Wooldridge, & Rosenschein, 2010) and sharing the gains from regional cooperation in the electricity market (Gately, 1974). Although power indices allow finding fair allocations of rewards and costs, they are susceptible to some forms of strategic behavior (Yokoo, Conitzer, Sandholm, Ohta, & Iwasaki, 2005; Aziz, Bachrach, Elkind, & Paterson, 2011; Zuckerman, Faliszewski, Bachrach, & Elkind, 2012).

The differences between the Banzhaf and Shapley-Shubik indices were analyzed (Straffin, 1977), and each index was shown to reflects specific conditions in a voting body. Different axiomatizations of these two indices have been proposed (Shapley, 1953; Dubey & Shapley, 1979; Lehrer, 1988; Straffin, 1988; Laruelle, 1999; Laruelle & Valenciano, 2001).

## 4.2 Cooperative Games Over Networks

Our model was based on a network goal of connecting a set of primary servers. Other cooperative games over networks have been proposed, modeling other goals in networks. Some related work studies one such model, which deals with a cost sharing mechanism for multicast transmissions (Feigenbaum et al., 2001; Moulin & Shenker, 2001). This body of related work focuses on a weakly budget-balanced implementation (where, as opposed to our model the total payments by the agents may exceed the total cost incurred), or on mechanisms that are only resistant to deviations by *single* agents. Yet another model examines buying a path between a source and a target in a network (Archer & Tardos, 2002). This is a mechanism design model which examines the problem of eliciting truthful reports from the edges regarding their true costs, in contrast to our work which focuses on a cooperative game with full information, and in which the agents incur no cost for allowing the use of their resources.

Another similar model considers a scenario where agents control edges in a network flow graph, and a coalition wins if it can maintain a certain required flow between a source and a target (Bachrach & Rosenschein, 2009). In that specific model finding the Banzhaf index of an edge in that domain is #P-complete, though there is a polynomial algorithm for some restricted cases. We handle a very different scenario where agents are required to maintain *connectivity*, rather than a certain flow. Also, we are interested in maintaining this connectivity between *every* two primary vertices, rather than two specific vertices (we can simulate the case of two specific servers by having only two primary servers). Also, in our work the agents are the servers in the communication network, rather than the links.

The model of Minimum Cost Spanning Tree Game (Bird, 1976; Granot & Huberman, 1981, 1984) (MCSTG) is also quite similar to our model. This is a cost sharing game, where the agents are the vertices in a complete edge-weighted graph, and the cost of a coalition is the minimal weight of a tree that connects all the coalition's vertices to a designated root $r$. More formally, the vertices of the graph include the agents $I = \{1, 2, \ldots, n\}$ and the designated root $r \notin I$, so $V = I \cup \{r\}$, and the graph is a complete edge weighted graph (with $n(n + 1)/2$ edges); the cost of a coalition $S$ is the weight of the minimal spanning tree on the subgraph induced by $S \cup \{r\}$. Granot et. al show that MCSTGs always have non empty cores (1981) and provide algorithms for finding core imputations or computing the nucleolus (1984). Our CG model is quite different—CGs are a revenue sharing game, where no costs are associated with edges, and where "backbone" vertices are allowed. Most notably, our CGs are simple games (where a coalition either wins or loses), and the core of our CGs can sometimes be empty.

A generalization of MCSTGs called Steiner Tree Games (Skorin-Kapov, 1995), STGs, is also somewhat similar to our model. STGs do allow nodes that are not players, similarly to our "backbone" vertices, and may sometimes have an empty core. Related work on this model discusses a certain sufficient, but not necessary, condition for the core on an STG to be non-empty, and a polynomial algorithm for testing that condition (Skorin-Kapov, 1995). Since the core may be non-empty even if that condition does not hold, this does not allow polynomially determining if the core of an STG is nonempty. Again, as opposed to our CGs, STGs are non-simple cost sharing games. Further, we also examine computational aspect of power indices and core-relaxations, rather than non-emptiness of the core that is the focus of that related work. Our results do show that in CGs we can determine whether the core is empty in polynomial time and even return a representation of the core.

Another twist on MCSTGs results in a model where the cost for a coalition $S$ is the weight of the minimal tree that spans the vertices in $S \cup \{r\}$ in the entire graph, rather than the graph induced by $S \cup \{r\}$ (Faigle, Kern, Fekete, & Hochstättler, 1997). In other words, it is allowed to use nonmembers of the coalition to connect members of the coalition to the root. In this model the core may be empty, and testing for core-emptiness is an NP-complete problem (Faigle et al., 1997). Further results regarding this model show that computing the nucleolus for this game is NP-hard (Faigle, Kern, & Kuipers, 1998).

The variations of the MCSTG discussed above (Granot & Huberman, 1981; Skorin-Kapov, 1995; Faigle et al., 1997) are cost sharing games. In such games, the coalition must achieve a certain goal and each agent is endowed with some resources (for example, the ability to use the edges adjacent to the agent). However, using each such resource is associated with a cost, and a coalition attempts to minimize the total cost it incurs.[12] It is possible to "convert" a cost-sharing game to a reward sharing game (Moulin, 2002; Chalkiadakis et al., 2012),[13] for example by defining the utility of a coalition to be some very high constant reward minus the cost the coalition incurs to achieve its goal in the original cost sharing game. However, our model crucially depends on the assumption that a node incurs no cost for allowing the use of its links, so all coalitions that achieve the network goal have the same utility. We believe this better characterizes domains such as a computer network that has *already been constructed*, where the links of a node are simply available for it to use. The MCSTG better models domains where agents must make decisions about which links to build *in the future* and where constructing a link requires an investment on behalf of the agents.

Yet another related network based cooperative games is the Spanning Connectivity Games (Aziz et al., 2009) (SCG for short). SCGs are similar to our CGs in that they are cooperative network reward sharing game. However, as opposed to our model, in SCGs the players are the *edges* of a multigraph, and a coalition wins if it manages to span all the nodes of the network. Yet another similar reward sharing game is the Path Disruption Games (Bachrach & Porat, 2010) where the coalition attempts to *disrupt* connectivity between two specific vertices. Although those domains are combinatorially different from CGs, this previous work examines similar solution concepts: the core and the Shapley value. For example, this related work shows that computing power indices in these domains is hard and that there are computationally tractable algorithms for solving core-related problems (at least in somewhat restricted domains).

---

12. In some such games a coalition may not even be able to achieve its goal at all, in which case we can define its cost to be infinite.

13. Sometimes reward sharing games are also called surplus sharing games.

### 4.2.1 COMPUTING POWER INDICES AND THE CORE

As their name suggests, power indices can also be though of as a measure of the significance of agents in a game. However, although both the Shapley and Banzhaf power indices are defined not only for voting games but for any simple cooperative game, relatively little work has examined the use of power indices for measuring the importance of players in non-voting scenarios. The complexity of computing power indices depends on the concrete representation of the game. When the game is defined only by the value of each coalition, in the form of an oracle that tests a certain coalition and answers whether it wins or loses, calculating power indices is difficult. A naive algorithm for calculating the power index of an agent $a_i$ enumerates all coalitions or permutations containing $a_i$. Since there are exponentially many such coalitions or permutations, the naive algorithm is exponential in the number of agents.

Related work discusses algorithms for calculating power indices in weighted majority games (Matsui & Matsui, 2000), and shows that calculating the Banzhaf and Shapley-Shubik indices in weighted voting games are both NP-complete problems (Matsui & Matsui, 2001). Since weighted voting games are a restricted case of simple coalitional games, the problem of calculating either index in a general coalitional game is of course NP-hard. In fact, in certain cases, calculating power indices is not just NP-hard but also #P-hard. Deng and Papadimitriou (1994) show that computing the Shapley-Shubik index in weighted voting games is #P-complete. Other research has derived hardness results for power indices in other game classes, such as Coalitional Skill Games (Bachrach & Rosenschein, 2008) which are based on a set-covering problem, or in a rule based cooperative game representation called the Multi-Attribute Coalitional Game language (Ieong & Shoham, 2006).

Our hardness results regarding computing the power indices might make using this concept seem less attractive. However, There are many results on comparing and approximating power indices, in general and in restricted domains (Owen, 1975; Deng & Papadimitriou, 1994; Conitzer & Sandholm, 2004; Bachrach, Markakis, Resnick, Procaccia, Rosenschein, & Saberi, 2010; Faliszewski & Hemaspaandra, 2009). This line of work shows that although computing power indices *exactly* is generally hard, *estimating* them with a high degree of accuracy is computationally tractable. For example, "problematic" vertices in a network can be tractably found by employing an approximation method (Bachrach et al., 2010) which can handle arbitrary cooperative games, so long as it is possible to compute the value of a coalition in polynomial time (which is easy to do in CGs). The algorithm provided by Bachrach et al. (2010) estimates the power indices and returns a result that is probably approximately correct: given a game in which a player's true power index is $\beta$, and given a target accuracy level $\epsilon$ and confidence level $\delta$, the algorithm returns an approximation $\hat{\beta}$ such that with probability at least $1 - \delta$ we have $|\beta - \hat{\beta}| \leq \epsilon$ (i.e., the result is approximately correct, and is within a distance $\epsilon$ of the correct value). Its running time is logarithmic in the confidence and quadratic in the accuracy, so the approach is tractable even for high accuracy and confidence. Methods for computing power indices were also examined in the context of games with uncertain agent failures (Bachrach, Meir, Feldman, & Tennenholtz, 2011; Bachrach, Kash, & Shah, 2012b).

While our treatment of the model is game theoretic, in network domains problems akin to calculating power indices can also be formulated as network reliability problems. The computational complexity of such problems has been studied in several papers. Classical network reliability problems consider an undirected graph $G = \langle V, E \rangle$, when each edge $e \in E$ has a probability assigned to it, $p_e$. This is the probability that edge $e$ remains in the surviving graph.

One prominent problem is that of s-t connectivity probability (STC-P): given the above domain, compute the probability of having a path between $s, t \in V$ in the surviving graph. Another prominent problem is that of full connectivity probability (FC-P): given the above domain, compute the probability that the surviving graph is connected (so that there is a path between any two vertices). One seminal paper by Valiant (1979a) proved that STC-P is #P-hard. Provan and Ball (1983) showed that FC-P is also #P-hard.

Some of the problems we study are similar to FC-P. For example computing the Banzhaf power index in CGs is a *very specific* case of FC-P, where the probability of every vertex subset is equal (or equivalently, where each vertex has a 50% probability of failures, and failures are independent). Since this is a restricted case, we cannot use the hardness result of Provan and Ball (1983), and have to prove that even the restricted case is #P-complete (which we did, in Section 3.1).

## 5. Conclusions

We have considered the computational aspects of reward sharing in a network connectivity scenario, and its applications to network reliability. We modeled a communication network as a simple coalitional game, and showed how various game-theoretic solution concepts can be used to characterize reasonable reward sharing agreements agents might make and to to find significant possible points of failure in the network. We have shown that in this domain, for general graphs, computing the Shapley and Banzhaf power indices is #P-complete. Despite this high complexity result for the general domain, we also gave a polynomial result for the restricted domain where the graph is a tree.

We have also shown that computing the core can be done in polynomial time in any CG, and gave a simple characterization of the instances when the core is non-empty in CGs. On the other hand, we have shown that in general CGs, testing if an imputation is in the $\epsilon$-core (or equivalently computing the maximal excess of a coalition under the imputation) is coNP-complete. We have also given a characterization of the core in tree CGs, and shown how testing for $\epsilon$-core imputations can be done in polynomial time for tree CGs.

It remains a topic of future research to tractably compute power indices in CGs over restricted domains. We also note that the Shapley and Banzhaf indices are not the only power indices studied in the literature, so studying computational aspects of other indices is also of interest. We have also examined the core, $\epsilon$-core and least core. Our hardness results show that computing the maximal excess of a coalition is computationally hard in general CGs. It would be interesting to see if it could be approximated, or exactly computed in restricted domains other than trees. Yet another open question is that of computing other game theoretic solution concepts in CGs or restricted CGs. One interesting problem is computing the nucleolus (Schmeidler, 1969) in CGs.[14] Another interesting direction is ex-amining coalition formation models (Dang, Dash, Rogers, & Jennings, 2006; Greco, Malizia, Palopoli, & Scarcello, 2011) and analyzing the coalition structure generation problem (Rahwan, Ramchurn, Dang, & Jennings, 2007; Ohta, Conitzer, Ichimura, Sakurai, Iwasaki, & Yokoo, 2009; Bachrach, Meir, Jung, & Kohli, 2010) in our domain.

---

14. For example, in tree CGs, the imputation which equally allocates all the rewards to the essential vertices $V_{es}$ is the nucleolus. However, we believe that there may even exist restricted CG domains where computing the $\epsilon$-core or the least core is tractable, but computing the nucleolus is hard.

## Acknowledgments

This work was partially supported by Israel Science Foundation grants #898/05 and #1227/12, and Israel Ministry of Science and Technology grant #3-6797.

A preliminary version of this paper was presented at the Seventh International Joint Conference on Autonomous Agents and Multiagent Systems (AAMAS 2008). This extended version includes a more complete analysis of power indices in CGs, complementing the previous results regarding the Banzhaf index with new results regarding the Shapley value. We have also expanded the work regarding stability based solutions, so this version contains an analysis of problems related to the $\epsilon$-core and least core, showing that such problems are coNP-hard in general CGs, but can be solved in polynomial time for the restricted case of CGs over trees. It also includes a more complete analysis of the core in CGs played over trees.